\documentclass[a4paper,fleqn]{cas-dc}
\newcommand{\std}[2]{#1\textsubscript{\scriptsize\,$\pm$\,#2}}

\usepackage[numbers]{natbib}
\usepackage{subcaption}
\usepackage{booktabs}
\usepackage{multirow}
\usepackage{tabularx}
\usepackage{array}
\usepackage{placeins}
\usepackage[ruled,vlined]{algorithm2e}
\usepackage[normalem]{ulem}

\def\tsc#1{\csdef{#1}{\textsc{\lowercase{#1}}\xspace}}
\tsc{WGM}
\tsc{QE}
\tsc{EP}
\tsc{PMS}
\tsc{BEC}
\tsc{DE}

\begin{document}
\let\WriteBookmarks\relax
\def\floatpagepagefraction{1}
\def\textpagefraction{.001}

\renewcommand{\topfraction}{0.90}
\renewcommand{\bottomfraction}{0.80}
\renewcommand{\topfraction}{0.95}
\renewcommand{\bottomfraction}{0.90}
\renewcommand{\textfraction}{0.05}
\shorttitle{APRIL}
\shortauthors{X.~Li et~al.}

\title [mode = title]{APRIL: Ambiguity-Aware Expert Routing for Respiratory Sound Domain-Incremental Learning}

\author[1]{Xinhong Li}[type=editor,
                        orcid=0000-0001-7063-2327]

\affiliation[1]{organization={School of Computer Science, Wuhan University},
                city={Wuhan},
                state={Hubei},
                country={China}}

\author[1]{Weiping Tu}
\cormark[1]
\ead{tuweiping@whu.edu.cn}

\cortext[cor1]{Corresponding author}

\begin{abstract}
Automated respiratory sound classification enables continuous, non-invasive respiratory assessment, yet long-term deployment remains challenging due to domain shifts from evolving recording devices and acquisition periods. When privacy and governance regulations prohibit replaying historical data, sequential learning easily overwrites previously acquired knowledge. Preserving dedicated domain experts prevents parameter interference, but selecting the correct expert at inference is difficult under overlapping acoustic distributions and absent domain labels. To resolve this dilemma, we propose APRIL, an exemplar-free domain-incremental framework for RSC. Built on a shared frozen BEATs backbone, APRIL sequentially inherits and isolates low-rank experts to enable new-domain adaptation without altering earlier models. A multi-level prototype router retrieves the top-two candidate experts and dynamically selects the final model by integrating domain-routing posteriors with classification confidence, adaptively weighted by matching ambiguity. We establish standardized device- and temporal-incremental benchmarks on the ICBHI and SPRSound datasets across forward and reversed domain orders. Evaluated over five random seeds, APRIL consistently achieves the highest final-session average Score among compared continual learning approaches, reaching 64.4\% and 65.4\% on ICBHI, and 83.1\% and 82.2\% on SPRSound under forward and reversed sequences, respectively. Crucially, average forgetting remains under 1.0\% across all protocols, demonstrating robust domain adaptation and long-term knowledge retention without data replay. Overall, APRIL offers a reliable and privacy-preserving paradigm for deploying continuous respiratory sound classification across real-world healthcare settings.
\end{abstract}


\maketitle

\section{Introduction}

Chronic respiratory diseases impose a substantial global burden of disability and premature mortality, underscoring the need for accessible respiratory assessment~\cite{CHEN2025128694,KHAN2021115456}. Automated respiratory sound classification (RSC) offers a low-cost, non-invasive means of analysing respiratory recordings~\cite{shridhar2026lung,11461974,sarkar2015auscultation}. Adventitious sounds, including crackles and wheezes, provide acoustic cues associated with respiratory abnormalities~\cite{WU2024121089,DELATORRECRUZ2020113212}. Recognizing these events automatically supports consistent analysis and provides complementary evidence for clinical examination.

Recent advances in RSC increasingly build on pretrained models such as AST, CLAP, and BEATs~\cite{bae23b_interspeech,kim24f_interspeech,jeong25_interspeech}. Their transferable representations provide a foundation for adaptations tailored to respiratory sound classification. In practice, however, respiratory recordings continue to accumulate as data collection proceeds, and recording devices and acquisition conditions can change over time. Data consequently become available in successive batches, potentially with different distributions, rather than being fully available at the outset~\cite{jiang2025advances}.

\begin{figure*}[t]
    \centering

    \begin{subfigure}[t]{0.49\textwidth}
        \centering
        \includegraphics[width=\linewidth]
        {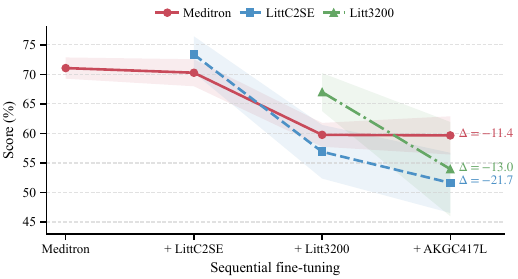}
        \caption{Device sequence on ICBHI.}
        \label{fig:forgetting_icbhi}
    \end{subfigure}
    \hfill
    \begin{subfigure}[t]{0.49\textwidth}
        \centering
        \includegraphics[width=\linewidth]
        {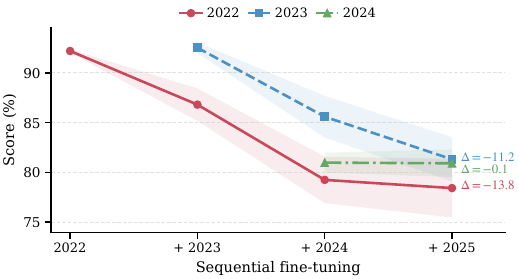}
        \caption{Temporal sequence on SPRSound.}
        \label{fig:forgetting_sprsound}
    \end{subfigure}

   \caption{Catastrophic forgetting under sequential fine-tuning. Each curve reports the Score on a previously learned domain after successive domain updates. Lines and shaded regions denote the mean and standard deviation over five seeds, respectively. $\Delta$ denotes the change between the stage at which a domain is first learned and the final stage.}
    \label{fig:sequential_forgetting}
\end{figure*}

This setting can be formulated as domain-incremental learning (DIL), in which new domains arrive sequentially while the classification label space remains fixed. For RSC, domains can correspond to recording devices or acquisition periods. Privacy and institutional data-governance requirements can restrict the retention or reuse of historical recordings~\cite{huang2026federated,birashk2025federated}. Without these recordings for replay, a model must adapt to incoming data while retaining its performance on earlier domains. Direct fine-tuning on the current domain can change representations needed for previously learned sounds, leading to catastrophic forgetting. Figure~\ref{fig:sequential_forgetting} illustrates this problem: sequential fine-tuning of BEATs produces substantial performance losses on earlier domains in both the ICBHI device sequence and the SPRSound temporal sequence.

Continual learning methods address forgetting through different forms of knowledge preservation. Exemplar-based replay retains historical examples for use during subsequent updates~\cite{xiao2022continual}, but requires continued access to earlier recordings. Without storing those recordings, knowledge distillation constrains changes in predictions and representations, while feature transformation compensates for changes in the feature space~\cite{mulimani2024class,chen2026aft}. In RSC, Lung-CL combines latent generative replay with spectrum-aware distillation~\cite{lai26_interspeech}. These approaches preserve earlier knowledge while adapting shared model components, but new-domain updates can still affect parameters needed for earlier domains. To reduce this interference, domain-specific approaches~\cite{mulimani2024online,10890481,park2026domain} introduce dedicated components for new domains while keeping previously learned components fixed. This isolates earlier parameters from subsequent updates, but introduces an inference challenge: selecting a suitable component when the recording's domain is unknown.

S-Prompts addresses this challenge through nearest-centroid routing, matching input features to domain cluster centers in a frozen representation space and activating the corresponding domain-specific module~\cite{wang2022s}. However, respiratory recordings collected in different periods can share similar acoustic characteristics. When their domain representations overlap, nearest-centroid routing can yield ambiguous assignments~\cite{feng2026hedp}. Moreover, domain identity alone does not determine which expert will classify a recording most effectively: experts trained on other domains may provide useful class-specific knowledge~\cite{li2025addressing}. An alternative to domain matching is therefore to select experts based on their predictions. ADIL adopts this approach, selecting the domain-specific branch with the lowest prediction entropy~\cite{10890481}. This brings classification evidence into expert selection, but requires evaluating every available branch. As an RSC system encounters more devices and acquisition periods, evaluating the growing expert bank for each recording increases inference computation.

Based on these considerations, we propose APRIL, an exemplar-free framework for domain-incremental respiratory sound classification. APRIL shares a frozen BEATs backbone across compact domain-specific LoRA adapters and classification heads. Each new expert inherits the adapter and head of the preceding expert before adaptation to the current domain, while earlier experts remain unchanged. At inference, APRIL separates domain matching from final expert selection. Compact multi-level prototypes represent each domain in a fixed reference space, keeping routing features independent of expert updates. Domain matching identifies a small set of candidate experts, whose predictions are then considered together with their domain relevance to select a single expert. This allows prediction-informed selection without evaluating the entire expert bank. APRIL requires no domain identity at inference and retains domain prototypes rather than historical recordings or sample-level feature caches.

Our contributions are summarized as follows:
\begin{itemize}
    \item We propose APRIL, an exemplar-free framework for
    respiratory sound DIL that combines sequential knowledge
    inheritance with parameter isolation. Inherited initialization
    reuses previously learned respiratory sound knowledge, while
    isolated LoRA--classifier experts allow new-domain adaptation
    without modifying earlier experts.

    \item We introduce an ambiguity-aware expert selection strategy
    that combines multi-level domain matching with classification
    confidence. By considering both routing and classification
    evidence, APRIL selects experts for respiratory sound recognition
    without requiring domain labels at inference.

    \item We establish complementary device- and temporal-incremental
    protocols on ICBHI and SPRSound, respectively. APRIL achieves
    the highest FSA among the compared continual-learning methods
    under both forward and reversed orders on both datasets,
    while maintaining low forgetting.
\end{itemize}

\section{Related Work}

Respiratory sound classification and domain-incremental learning provide two complementary foundations for this study. We first review acoustic representations and learning architectures for respiratory sound classification, including recent adaptations of pretrained models. We then discuss approaches to preserving earlier knowledge during sequential learning, with particular attention to domain-specific components and their selection at inference.

\subsection{Respiratory Sound Classification}

Public datasets, including ICBHI~\cite{rocha2017alpha} and SPRSound~\cite{zhang2022sprsound}, have supported the development and comparison of respiratory sound classification methods.

\textit{(1) Handcrafted representations.}
Early approaches extracted explicitly designed acoustic descriptors, such as mel-frequency cepstral coefficients (MFCCs), linear predictive coefficients, and median frequency. These descriptors were combined with statistical classifiers, including Gaussian mixture models, hidden Markov models, and support vector machines~\cite{haider2019respiratory,jakovljevic2017hidden}. In these systems, feature extraction was designed separately from classifier training, making the selection of informative acoustic descriptors a central part of model development.

\textit{(2) Deep neural architectures.}
With advances in deep learning, RSC increasingly shifted toward learning acoustic representations from waveforms or time--frequency inputs. CNN- and ResNet-based models have been widely investigated~\cite{gairola2021respirenet,li2021lungattn}. Multi-branch architectures extend this approach by learning and fusing complementary feature streams rather than relying on a single input representation~\cite{ma2019lungbrn,xu2021arsc}. Other studies incorporate recurrent networks or temporal convolutional modules to model dependencies across a respiratory recording~\cite{acharya2020deep,le2025respiratory}. These architectures emphasize the joint modeling of local acoustic patterns and their temporal context.

\textit{(3) Pretrained models and domain adaptation.}
Recent studies increasingly adapt pretrained models to RSC, using representations learned from larger datasets as the foundation for task-specific learning~\cite{shi2023lung,toikkanen25_interspeech, dong2025respiratory}. For example, Bae et al. combine AST with Patch-Mix augmentation and contrastive learning~\cite{bae2023patch}, while Jeong and Kim introduce patient-aware feature alignment on BEATs to address inter-patient variability~\cite{jeong25_interspeech}. Multimodal approaches also incorporate patient and recording context: BTS adapts CLAP using respiratory audio and textual metadata~\cite{kim24f_interspeech},
whereas TIAMO integrates metadata-derived text,
spectrogram images, and raw audio within a tri-modal
framework~\cite{le2026tiamo}.

Recording variability has also motivated
stethoscope- and metadata-guided contrastive approaches
to domain adaptation~\cite{kim2024stethoscope,
kim2025adaptive}.
Meanwhile, parameter-efficient methods use adapters
or low-rank updates to transfer pretrained models
to RSC with fewer trainable
parameters~\cite{xiao24_interspeech,dong2025low}.
These studies address how pretrained representations
can be adapted to respiratory sounds and heterogeneous
recording conditions.

Very recently, Lung-CL extends RSC to buffer-free
domain-incremental learning through class-specific
generative latent replay and spectrum-aware
distillation~\cite{lai26_interspeech}.
It preserves earlier knowledge within a continually
updated shared model under shifts across heterogeneous
respiratory sound datasets.
Our work studies complementary protocols defined by
individual recording devices and annual releases,
focusing on the preservation and selection of
domain-specific experts.

\subsection{Domain-Incremental Learning}

Continual learning investigates how models acquire new
knowledge while retaining what they have previously
learned. We review the main strategies underlying
domain-incremental learning and their applications
to audio classification.

\textit{(1) Rehearsal-based methods.}
Rehearsal preserves earlier knowledge by reintroducing
historical information during subsequent training.
ER mixes buffered examples with current
data~\cite{rolnick2019experience}, while DER++
additionally matches stored model outputs to preserve
earlier predictions~\cite{buzzega2020dark}.
Generative replay provides an alternative by synthesizing
examples from previously learned distributions instead
of retaining the original
observations~\cite{shin2017continual}.
These approaches provide supervision from earlier tasks,
but exemplar replay requires continued access to stored
data, whereas generative replay relies on maintaining
a generator that adequately represents past distributions.

\textit{(2) Regularization-based methods.}
To reduce reliance on stored historical examples,
regularization-based methods constrain changes to
a shared model during sequential training.
EWC penalizes changes to parameters important for
earlier tasks~\cite{kirkpatrick2017overcoming},
whereas LwF uses knowledge distillation to preserve
earlier predictions~\cite{li2017learning}.
Beyond parameter values and predictions, NEig-OWM
incorporates null-space information into orthogonal
weight modification to guide subsequent
updates~\cite{liao2025continuous}.
Such constraints must balance knowledge retention
against adaptation: strong constraints can restrict
learning under substantial distribution shifts,
while weaker constraints leave earlier knowledge
more exposed to interference~\cite{van2024continual}.

\textit{(3) Parameter-isolation methods.}
Parameter isolation addresses this conflict by
separating the parameters used for new learning
from those preserving earlier knowledge.
Network expansion, pruning, and parameter masking
provide dedicated capacity while protecting previously
learned parameters~\cite{rusu2016progressive,
mallya2018packnet}.
With pretrained backbones, prompts and adapters
offer more compact implementations.
S-Prompts learns independent domain prompts and
uses feature-based matching to select them at
inference~\cite{wang2022s}.
PINA combines domain-specific alignment modules
with a shared classifier, allowing incoming domains
to be aligned with previously learned class
concepts~\cite{wang2024non}.

LoRA-based continual learning further controls
interference through the structure of low-rank updates.
InfLoRA constrains new-task updates to subspaces
that preserve earlier knowledge~\cite{liang2024inflora}.
BiLoRA addresses the difficulty of maintaining
strictly orthogonal task subspaces by constructing
almost-orthogonal spaces through fixed orthogonal
bases~\cite{zhu2025bilora}.
To support knowledge sharing across domains,
ISPSL combines shared and domain-specific low-rank
components rather than separating all updates
completely~\cite{wang2024importance}.
For methods that retain multiple domain-specific
modules, preserving their parameters must also
be accompanied by an appropriate selection mechanism
when domain identities are unavailable at inference.

\textit{(4) Audio continual and domain-incremental learning.}
Recent audio studies investigate continual learning
through acoustic feature preservation and adaptation
of pretrained representations.
AFT transforms earlier-class temporal features
to account for changes in the representation
space during exemplar-free class-incremental
learning~\cite{chen2026aft}.
PACE combines analytic classification with
subspace-constrained parameter-efficient adaptation
for continual learning with pretrained audio
models~\cite{li2026pace}.
For domain-incremental audio classification,
ADIL separates domain-shared and domain-specific
parameters. When domain labels are unavailable,
it evaluates the available branches and selects
the prediction with the lowest
entropy~\cite{10890481}.
These studies provide a basis for continual adaptation
to changing acoustic conditions.
For RSC, this involves learning from recordings acquired
across different devices and collection periods.
Domain-specific models offer a way to preserve earlier
knowledge, but their use also requires selecting an
appropriate expert when the test domain is unknown.

\section{Method}
\label{sec:method}
\begin{figure*}[t]
    \centering
    \includegraphics[width=\textwidth]{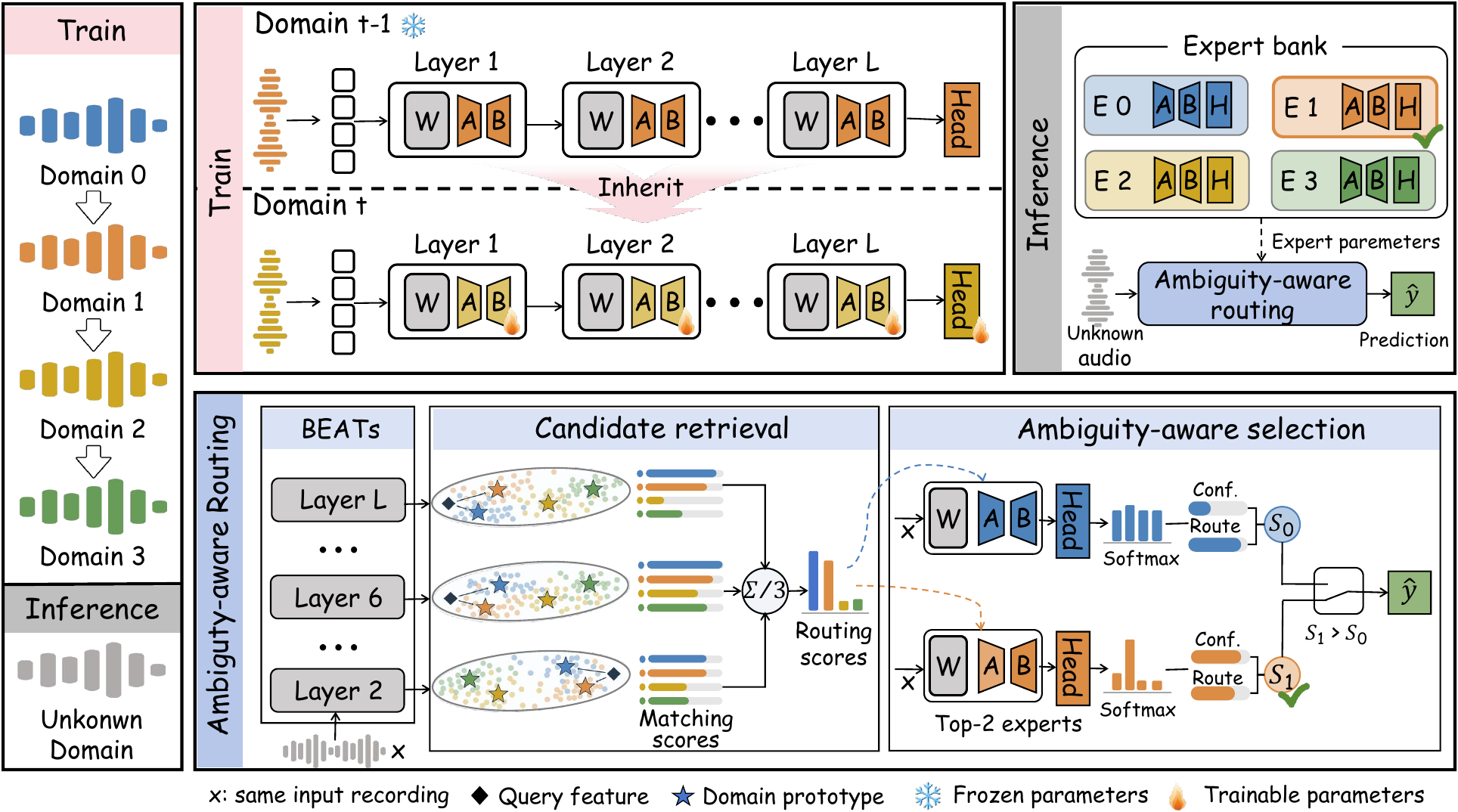}
    \caption{Overview of APRIL.
    Each new domain expert inherits the preceding expert's LoRA
    adapter and classification head, while the shared backbone and
    earlier experts remain frozen.
    During inference, multi-level prototype matching retrieves two
    candidate experts, and routing posterior and classifier confidence
    jointly determine which candidate's prediction is returned.}
    \label{fig:april_framework}
\end{figure*}

APRIL addresses respiratory sound DIL through isolated
domain adaptation and ambiguity-aware expert selection,
without replaying historical recordings.
As illustrated in Fig.~\ref{fig:april_framework},
new experts inherit the preceding expert's LoRA adapter
and classifier before adaptation, while earlier experts
remain unchanged.
During inference, multi-level prototype matching retrieves
two candidate experts in a frozen reference space.
Their routing posteriors and classification confidence
jointly determine which expert provides the final prediction.

\subsection{Problem formulation}
\label{sec:problem_formulation} 
Let $\{\mathcal{D}_t\}_{t=0}^{T-1}$ denote a sequence of respiratory-sound domains, where each domain contains a training subset $\mathcal{D}_t^{\mathrm{tr}}$ and a held-out test subset $\mathcal{D}_t^{\mathrm{te}}$. All domains within a benchmark share the same label space, while their data distributions change sequentially. At session $t$, the model is updated using only $\mathcal{D}_t^{\mathrm{tr}}$, without access to the training data of previous domains. After each session, it is evaluated separately on $\mathcal{D}_j^{\mathrm{te}}$ for every observed domain $j \leq t$. Domain identities are available during training but are not provided during inference.

\subsection{Pretrained audio backbone and low-rank adaptation}
\label{sec:backbone_adaptation}

APRIL uses the pretrained BEATs encoder as its shared
audio backbone~\cite{pmlr-v202-chen23ag}.
BEATs learns audio representations through acoustic
token prediction.
Its frontend converts a recording into normalized
log-Mel filterbank features, which are divided into
non-overlapping time--frequency patches and projected
into a token sequence.
Successive Transformer blocks process these tokens,
providing features at different encoder depths.

We introduce domain-specific adaptation through
LoRA~\cite{hu2022lora} in the key and value projections
of every self-attention block.
Let $\theta$ denote the frozen backbone parameters.
For a pretrained projection matrix
$W\in\mathbb{R}^{d_{\mathrm{out}}\times d_{\mathrm{in}}}$,
expert $t$ uses
\begin{equation}
    \widetilde{W}_t=W+sB_tA_t,
    \label{eq:lora_projection}
\end{equation}
where $A_t\in\mathbb{R}^{r\times d_{\mathrm{in}}}$
and $B_t\in\mathbb{R}^{d_{\mathrm{out}}\times r}$
are trainable low-rank factors, $r$ is the adapter
rank, and $s$ is a scaling factor.
Each adapted projection has its own pair of factors.
Together, these factors constitute the domain-specific
adapter parameters $\phi_t$.

For a recording $x$, the adapted encoder produces
the final token sequence $F_{\theta,\phi_t}(x)$.
Its mean-pooled representation is passed to a
domain-specific linear classifier:
\begin{equation}
    \begin{aligned}
        \mathbf{h}_t(x)
        &=\operatorname{MeanPool}
          \left(F_{\theta,\phi_t}(x)\right),\\
        \mathbf{p}_t(x)
        &=\operatorname{softmax}
          \left(U_t\mathbf{h}_t(x)+\mathbf{b}_t\right).
    \end{aligned}
    \label{eq:expert_prediction}
\end{equation}
Here, $U_t\in\mathbb{R}^{C\times d}$ and
$\mathbf{b}_t\in\mathbb{R}^{C}$ form the classifier
parameters $\psi_t$, where $d$ is the encoder feature
dimension and $C$ is the number of respiratory
sound classes.

For prototype construction and domain matching,
the same backbone serves as a reference encoder
with the LoRA adapters disabled.
Its intermediate and final features provide a
common reference space that remains unchanged
as new experts are learned.
Expert predictions instead use the corresponding
domain's adapter and classifier.

\subsection{Sequentially inherited and isolated experts}
\label{sec:expert_learning}

APRIL builds its expert bank as domains arrive.
At session $t$, a new expert receives its own
LoRA adapter $\phi_t$ and classifier $\psi_t$.
All experts share the pretrained BEATs backbone,
while their domain-specific parameters are learned
from recordings of their respective domains.
This allows each expert to specialize in the
recording conditions of its assigned domain.

For the first domain, we randomly initialize the
$A$ matrices and set the $B$ matrices to zero,
following the initialization pattern of
LoRA~\cite{hu2022lora}.
The initial low-rank updates are therefore zero,
so the encoder initially retains its pretrained behavior.
The first classifier is also randomly initialized.

For subsequent domains, each new expert inherits
the adapter and classifier jointly trained on
the preceding domain.
Although recording devices or collection periods
may differ, all domains require recognition of
the same respiratory sound classes.
The inherited parameters therefore provide a
starting point already trained for this task.
Each new expert receives an independent copy
of these parameters, allowing it to adapt to
current-domain recordings separately from
the preceding expert.
Inheritance thus reuses previously learned
respiratory sound patterns while leaving room
for domain-specific adaptation.

At session $t$, we train the current expert on
$\mathcal{D}_t^{\mathrm{tr}}$ using the following
cross-entropy objective:
\begin{equation}
    \mathcal{L}_t
    =
    -\frac{1}{|\mathcal{D}_t^{\mathrm{tr}}|}
    \sum_{(x,y)\in\mathcal{D}_t^{\mathrm{tr}}}
    \log p_t(y\mid x),
    \label{eq:expert_loss}
\end{equation}
where $p_t(y\mid x)$ is the probability assigned
to the ground-truth class $y$ by the current expert.
Only $\phi_t$ and $\psi_t$ are updated.
The backbone and previously trained experts remain
frozen, so current-domain learning changes neither
the shared pretrained weights nor earlier experts.

After training, the current expert is frozen
and added to the bank.
Its learned parameters initialize the next expert,
while it remains available for subsequent inference.
Repeating this procedure expands the bank by one
expert per domain.
Sequential inheritance connects their learning
across sessions, while parameter isolation preserves
each expert after adaptation.

\subsection{Multi-level prototype-based candidate retrieval}
\label{sec:candidate_retrieval}

APRIL retrieves candidate experts by comparing the input
recording with prototypes from previously observed domains.
For this purpose, we use the shared frozen BEATs backbone
with its LoRA adapters disabled, referring to it as
the reference encoder.
A single forward pass provides intermediate and final
token features.
We mean-pool the tokens at each selected level
$\ell\in\mathcal{S}$ to obtain a recording-level
representation $\mathbf{z}^{(\ell)}(x)$.

For each incoming domain $t$, we apply cosine
(spherical) $K$-means~\cite{dhillon2001concept}
independently to its training features at each level.
Clustering uses all recordings in
$\mathcal{D}_t^{\mathrm{tr}}$, without their respiratory
sound labels.
The centers are initialized using farthest-point selection.
The algorithm then groups normalized features by cosine
similarity and updates each center to the normalized
mean of its cluster.
The resulting centers form the domain prototypes
$\{\boldsymbol{\mu}_{t,k}^{(\ell)}\}_{k=1}^{K}$,
with $K=2$ in the main configuration.
These prototypes are stored alongside their domain
association, while prototypes from earlier domains
remain unchanged.

At inference, the input recording is encoded in the
same way.
For each observed domain $j$, we compare its feature
at level $\ell$ with that domain's $K$ prototypes.
The highest cosine similarity is retained as the
level-wise matching score:
\begin{equation}
    m_j^{(\ell)}(x)
    =
    \max_{1\leq k\leq K}
    \operatorname{cos}
    \left(
        \mathbf{z}^{(\ell)}(x),
        \boldsymbol{\mu}_{j,k}^{(\ell)}
    \right),
    \label{eq:layer_matching}
\end{equation}
where $\operatorname{cos}(\cdot,\cdot)$ denotes
cosine similarity.
Thus, with $K=2$, each domain provides two similarities
at each level, and the larger one becomes its
matching score.

We then average the matching scores across the
selected levels to obtain one routing score
for each domain:
\begin{equation}
    d_j(x)
    =
    \frac{1}{|\mathcal{S}|}
    \sum_{\ell\in\mathcal{S}}m_j^{(\ell)}(x).
    \label{eq:domain_matching}
\end{equation}
These two steps correspond to the matching scores
and routing scores in Fig.~\ref{fig:april_framework}.
Each layer first compares the recording with the
domain prototypes; averaging then combines the
intermediate- and final-layer comparisons.
The highest routing scores identify up to two
candidate experts.

The routing scores also define a posterior over
all domains observed by session $t$:
\begin{equation}
    q_j(x)
    =
    \frac{\exp\left(d_j(x)/\tau\right)}
    {\sum_{u=0}^{t}\exp\left(d_u(x)/\tau\right)},
    \label{eq:routing_posterior}
\end{equation}
where $\tau$ is the routing temperature.
The next stage combines these routing posteriors
with the candidate experts' classification confidence
to select the final prediction.

\subsection{Ambiguity-aware expert selection}
\label{sec:expert_selection}

Recordings from different domains can share similar
acoustic characteristics.
When prototype matching gives similar support to
two domains, APRIL also considers the candidate
experts' classification confidence to select
the final prediction.

Let $e_1$ and $e_2$ denote the two retrieved experts,
ordered such that $q_{e_1}(x)\geq q_{e_2}(x)$.
Each candidate $j\in\{e_1,e_2\}$ processes the same
recording $x$ using its own adapter and classifier.
For $C$ respiratory sound classes, we define its
classification confidence using normalized
prediction entropy:
\begin{equation}
    \gamma_j(x)
    =
    1-
    \frac{\mathcal{H}\left(\mathbf{p}_j(x)\right)}
         {\log C},
    \label{eq:expert_confidence}
\end{equation}
where
$\mathcal{H}(\mathbf{p}_j(x))
=-\sum_{c=1}^{C}p_j(c\mid x)\log p_j(c\mid x)$.
Confidence is higher when the expert assigns most
of its probability to one class and is zero when
all classes receive equal probability.
Dividing by $\log C$ keeps the confidence score
between zero and one for different class counts.

We give classification confidence greater weight
when domain matching cannot clearly distinguish
the two candidates.
Conversely, a larger routing-posterior gap gives
more weight to domain matching.
Using the posteriors from
Eq.~\eqref{eq:routing_posterior}, we quantify
domain-matching ambiguity as
\begin{equation}
    a(x)
    =
    1-\left[q_{e_1}(x)-q_{e_2}(x)\right].
    \label{eq:routing_ambiguity}
\end{equation}
A smaller posterior gap produces a larger $a(x)$,
which increases the weight assigned to
classification confidence in the selection score below.

For each candidate $j\in\{e_1,e_2\}$, we normalize
its routing posterior within the retrieved pair:
\begin{equation}
    \widetilde{q}_j(x)
=q_j(x)/(q_{e_1}(x)+q_{e_2}(x)).
\end{equation}
The final selection score combines domain matching
and classification confidence:
\begin{equation}
    S_j(x)
    =
    \log\widetilde{q}_j(x)
    +
    \lambda\,a(x)\,\gamma_j(x),
    \label{eq:expert_selection_score}
\end{equation}
where $\lambda$ sets the base weight of
classification confidence.
The first term favors the candidate with the
higher routing posterior.
The second adds its classification confidence
with a recording-specific weight $\lambda a(x)$.
When the routing posteriors are close, this weight
increases, allowing the more confident expert
to have greater influence on the choice.
The second-ranked candidate can therefore be
selected if its higher classification confidence
compensates for its lower routing posterior.

We select the expert with the higher score
and return its most probable class:
\begin{equation}
    \begin{aligned}
        j^*
        &=\operatorname*{arg\,max}_{j\in\{e_1,e_2\}}
          S_j(x),\\
        \hat{y}
        &=\operatorname*{arg\,max}_{c\in\{1,\ldots,C\}}
          p_{j^*}(c\mid x).
    \end{aligned}
    \label{eq:final_prediction}
\end{equation}
The candidate experts are evaluated sequentially
on the shared backbone.
Inference uses one reference encoding followed
by two candidate evaluations and the final selection.
When only the first domain has been learned,
its expert provides the prediction directly.

\section{Experimental Setup}

\subsection{Datasets and domain-incremental protocols} 
\label{sec:datasets}

\begin{table*}[t]
\centering
\caption{Summary of the domain-incremental protocols. Each entry is
reported as training/test count. Patient totals count unique subjects,
and the global training and test patient sets are disjoint.}
\label{tab:dataset_protocols}
\small
\setlength{\tabcolsep}{5pt}
\renewcommand{\arraystretch}{1.10}

\begin{subtable}[t]{0.48\textwidth}
\centering
\caption{ICBHI device domains.}
\label{tab:icbhi_protocol_summary}
\begin{tabular}{@{}llccc@{}}
\toprule
Session & Device
& \shortstack{Recordings\\(Tr/Te)}
& \shortstack{Patients\\(Tr/Te)}
& \shortstack{Cycles\\(Tr/Te)} \\
\midrule
S0 & Meditron  & 77/50   & 41/23 & 895/561 \\
S1 & LittC2SE  & 50/37   & 13/10 & 348/246 \\
S2 & Litt3200  & 33/27   & 6/5   & 285/217 \\
S3 & AKGC417L  & 379/267 & 19/13 & 2,620/1,726 \\
\midrule
Total & -- & 539/381 & 76/50 & 4,148/2,750 \\
\bottomrule
\end{tabular}
\end{subtable}
\hfill
\begin{subtable}[t]{0.48\textwidth}
\centering
\caption{SPRSound temporal domains.}
\label{tab:sprsound_protocol_summary}
\begin{tabular}{@{}llccc@{}}
\toprule
Session & Year
& \shortstack{Recordings\\(Tr/Te)}
& \shortstack{Patients\\(Tr/Te)}
& \shortstack{Events\\(Tr/Te)} \\
\midrule
S0 & 2022 & 1,828/855 & 195/97 & 6,161/2,928 \\
S1 & 2023 & 481/343   & 54/38  & 1,878/1,113 \\
S2 & 2024 & 927/671   & 185/138 & 4,120/3,083 \\
S3 & 2025 & 647/425   & 87/74  & 2,302/1,608 \\
\midrule
Total & -- & 3,883/2,294 & 521/347 & 14,461/8,732 \\
\bottomrule
\end{tabular}
\end{subtable}
\end{table*}

We evaluate all methods on two complementary domain-incremental
respiratory sound classification protocols constructed from
ICBHI 2017~\cite{rocha2017alpha} and the 2022--2025 releases
of SPRSound~\cite{zhang2022sprsound}. ICBHI examines sequential
adaptation across recording devices, whereas SPRSound considers
domains defined by annual releases.
Table~\ref{tab:dataset_protocols} summarizes the domain-wise
statistics and training/test partitions. Both datasets use
fixed patient-disjoint splits, and all compared methods
receive identical data partitions and domain sequences.

\subsubsection{ICBHI 2017}

The ICBHI 2017 database contains 920 recordings from
126 patients, comprising 6,898 respiratory cycles labeled
as Normal, Crackle, Wheeze, or Both.
We define four device domains and introduce them in the
forward order Meditron $\rightarrow$ LittC2SE
$\rightarrow$ Litt3200 $\rightarrow$ AKGC417L.
The four-class label space remains unchanged throughout
the sequence.

We construct a fixed split of 76 training and 50 test
patients, corresponding to 539/381 recordings and
4,148/2,750 respiratory cycles in the training/test subsets.
Patient disjointness is enforced globally across devices:
all recordings from a patient belong to the same subset,
including patients recorded using multiple devices.
The sessions differ substantially in size.
In particular, AKGC417L contributes approximately 63\%
of all respiratory cycles, whereas the other devices
provide considerably smaller domains.
The protocol therefore evaluates sequential adaptation
across recording devices with unequal amounts of
training and test data.

\subsubsection{SPRSound 2022--2025}

The four annual SPRSound releases define the forward
temporal sequence 2022 $\rightarrow$ 2023
$\rightarrow$ 2024 $\rightarrow$ 2025.
We use a binary classification task in which Normal
retains its original label, while Rhonchi, Wheeze,
Stridor, Coarse Crackle, Fine Crackle, and
Wheeze+Crackle are grouped into Adventitious.
This common label space is maintained across all
four annual domains.

After excluding recordings without reliable patient
identifiers, we retain 6,177 recordings containing
23,193 annotated respiratory events.
The fixed patient-disjoint split includes 521 training
and 347 test patients, corresponding to 3,883/2,294
recordings and 14,461/8,732 events in the training/test
subsets.
For 2023--2025, we reconstruct the sessions from publicly
released recordings and event annotations using our
patient-disjoint partitions rather than adopting the
official challenge train/test partitions.
The annual sessions vary in size and class composition,
providing a temporal-incremental setting complementary
to the device-incremental protocol on ICBHI.

\paragraph{Domain-order protocols.}
The primary experiments follow the forward sequences
specified above and summarized in
Table~\ref{tab:dataset_protocols}.
For both datasets, we additionally evaluate the exact
reversed sequence to examine sensitivity to domain order.
The patient partitions, label spaces, and samples within
each domain remain fixed; only the arrival order changes.
After each session, the model is evaluated on the held-out
test subsets of all domains encountered so far.
Together, the forward and reversed experiments assess
adaptation and retention under four dataset/order
combinations.

\subsection{Compared methods}
\label{sec:compared_methods}

We compare APRIL with representative continual-learning methods
covering regularization, rehearsal, parameter-efficient adaptation,
and domain-incremental learning.

\textbf{Reference protocols.}
The \emph{Seq.~FT} sequentially fine-tunes a shared model without
an explicit forgetting-mitigation mechanism. The \emph{Joint Train.}
performs offline joint training using data from all domains. It is
order invariant and serves as an offline reference rather than a
continual-learning method.

\textbf{Conventional continual learning.}
LwF~\cite{li2017learning} preserves previous predictions through
knowledge distillation, whereas EWC~\cite{kirkpatrick2017overcoming}
penalizes changes to parameters estimated to be important for earlier
domains. ER~\cite{rolnick2019experience} replays samples retained in a
bounded memory, while DER++~\cite{buzzega2020dark} additionally
distills the historical logits associated with replayed samples.

\textbf{Parameter-efficient continual learning.}
InfLoRA~\cite{liang2024inflora} constrains low-rank updates using
subspaces derived from previous tasks. BiLoRA~\cite{zhu2025bilora}
constructs approximately orthogonal low-rank parameter spaces to
reduce interference, and ContLoRA~\cite{lu2026continual} progressively
integrates low-rank updates through cumulative unified optimization.

\textbf{Domain-incremental learning.}
DARE~\cite{jeeveswaran2024gradual} combines representation replay with
a staged optimization procedure. DUCT~\cite{zhou2025dual} consolidates
domain-specific representations and classifiers, whereas
SOYO~\cite{wang2025boosting} models domain characteristics to select
and combine domain-specific parameters.

\textbf{Respiratory-sound continual learning.}
Lung-CL~\cite{lai26_interspeech} combines class-conditional latent
replay with spectrum-aware knowledge distillation for respiratory-sound
continual learning. As executable training code was unavailable when
our experiments were conducted, we reimplemented its continual-learning
objectives from the published description and adapted them to the
common BEATs backbone.

All methods use identical patient-disjoint partitions, forward and
reversed domain orders, label spaces, and evaluation metrics.
The same BEATs initialization is used whenever architecturally
applicable. Method-specific components are preserved, and all
implementation adaptations and hyperparameter settings are reported
in the supplementary material.

\subsection{Implementation details}
\label{sec:implementation}

All respiratory events were resampled to 16~kHz and
standardized to 5~s by truncation or repetition padding,
without waveform or spectrogram augmentation.
APRIL used the AudioSet-pretrained BEATs model
with a frozen backbone.
Each domain had an independent linear classifier and
rank-16 LoRA adapters on the key and value projections
of every self-attention block, with scaling factor 1.0.

Experts were trained for 100 epochs using cross-entropy
loss and SGD with momentum 0.9 and zero weight decay.
The learning rates for LoRA parameters and classifiers
were $5\times10^{-3}$ and $1\times10^{-2}$, respectively,
and were reduced tenfold after epochs 60 and 80.
Training and evaluation batch sizes were 16 and 32.
The routing temperature and confidence weight were
0.002 and 1.0, respectively.
Experiments used five seeds, $\{1,2,3,4,5\}$,
with one NVIDIA RTX~4090 GPU per run.

\subsection{Evaluation metrics}
\label{sec:evaluation_metrics}

We evaluate respiratory sound classification performance using
specificity ($S_p$), sensitivity ($S_e$), and their arithmetic
mean, denoted as Score.
Let $N_x$ and $C_x$ denote the total number of samples and the
number of correctly classified samples in class $x$, respectively.
The subscripts $n$, $c$, $w$, $b$, and $a$ represent Normal,
Crackle, Wheeze, Both, and Adventitious, respectively.
Specificity is defined as $S_p=C_n/N_n$.
Sensitivity is computed as
$S_e=(C_c+C_w+C_b)/(N_c+N_w+N_b)$ for ICBHI and
$S_e=C_a/N_a$ for SPRSound, measuring the aggregate recognition
rate of adventitious sounds in each dataset.
The classification Score is defined as
$\mathrm{Score}=(S_p+S_e)/2$.
All classification measures are reported as percentages.

Following the performance-matrix evaluation framework commonly used in
medical continual learning~\cite{qazi2026continual}, let
$\mathbf{R}\in\mathbb{R}^{T\times T}$ denote the evaluation matrix,
where $R_{i,j}$ is the Score on domain $j$ after sequential training
through session $i$, with $1\leq j\leq i\leq T$.

The final-session average (FSA) measures the mean domain-wise performance
after completing the learning sequence:
\begin{equation}
\mathrm{FSA}
=
\frac{1}{T}
\sum_{j=1}^{T} R_{T,j}.
\label{eq:fsa}
\end{equation}

Average forgetting (AF) measures the decline from the best Score
observed for each previous domain to its final Score:
\begin{equation}
\mathrm{AF}
=
\frac{1}{T-1}
\sum_{j=1}^{T-1}
\left(
\max_{j\leq i\leq T} R_{i,j}
-
R_{T,j}
\right).
\label{eq:average_forgetting}
\end{equation}
The final session is included in the maximization, consistent with our
implementation, so that AF remains non-negative and equals zero when
the final Score is the best observed performance.

Backward transfer (BWT) compares the final performance on each previous
domain with its performance immediately after being learned:
\begin{equation}
\mathrm{BWT}
=
\frac{1}{T-1}
\sum_{j=1}^{T-1}
\left(
R_{T,j}-R_{j,j}
\right).
\label{eq:bwt}
\end{equation}
A negative BWT indicates forgetting, a value close to zero indicates
retention, and a positive value indicates beneficial backward transfer.
AF and BWT use different reference points: AF uses the best historical
performance, whereas BWT uses the performance obtained immediately
after learning the corresponding domain.

We further report the overall score (OS). After the final session,
the test sets of all domains are combined, and OS is calculated
using the Score definition given above.
Thus, FSA gives each domain equal weight by averaging the final
domain-wise Scores, whereas OS is computed from all test samples
together and is influenced by the number of samples contributed
by each domain.

\section{Results}
\subsection{Main comparison}
\label{sec:main_comparison}

\begin{table*}[p]
\centering
\footnotesize
\setlength{\tabcolsep}{4pt}
\renewcommand{\arraystretch}{1.10}

\caption{Performance comparison on ICBHI under forward and
reversed device orders. Results are mean $\pm$ standard deviation
over five seeds. Bold and underlined means indicate the best
and second-best results among continual-learning methods,
respectively. Seq.~FT and Joint Train. denote sequential
fine-tuning and joint training, respectively.}
\label{tab:icbhi_main_comparison}

\begin{tabularx}{\textwidth}{
    @{}l
    @{\hspace{3pt}}c
    *{8}{>{\centering\arraybackslash}X}
    @{}
}
\toprule
\multirow{2}{*}{\textbf{Method}}
& \multirow{2}{*}{\textbf{Year}}
& \multicolumn{4}{c}{\textbf{Forward order}}
& \multicolumn{4}{c}{\textbf{Reversed order}} \\
\cmidrule(lr){3-6}
\cmidrule(lr){7-10}
& 
& \textbf{FSA} ($\uparrow$)
& \textbf{AF} ($\downarrow$)
& \textbf{BWT} ($\uparrow$)
& \textbf{OS} ($\uparrow$)
& \textbf{FSA} ($\uparrow$)
& \textbf{AF} ($\downarrow$)
& \textbf{BWT} ($\uparrow$)
& \textbf{OS} ($\uparrow$) \\
\midrule

LwF
& 2016
& \std{\underline{62.58}}{2.13}
& \std{6.97}{3.69}
& \std{-6.36}{3.97}
& \std{59.98}{1.60}
& \std{\underline{65.19}}{0.91}
& \std{2.60}{1.43}
& \std{-0.30}{2.50}
& \std{60.21}{0.54} \\

EWC
& 2017
& \std{55.63}{1.54}
& \std{14.20}{2.25}
& \std{-14.20}{2.25}
& \std{57.04}{1.11}
& \std{64.79}{1.56}
& \std{4.53}{2.26}
& \std{-3.69}{3.47}
& \std{59.43}{0.41} \\

ER
& 2019
& \std{58.96}{2.31}
& \std{11.16}{2.60}
& \std{-11.09}{2.49}
& \std{60.18}{1.06}
& \std{64.41}{1.18}
& \std{3.25}{1.53}
& \std{-2.71}{1.49}
& \std{61.02}{0.90} \\

DER++
& 2020
& \std{56.68}{2.12}
& \std{12.60}{2.20}
& \std{-12.22}{2.68}
& \std{58.69}{0.61}
& \std{64.19}{1.89}
& \std{2.81}{1.69}
& \std{-2.21}{1.98}
& \std{60.72}{1.91} \\

InfLoRA
& 2024
& \std{59.39}{2.41}
& \std{6.12}{4.77}
& \std{-5.55}{4.46}
& \std{57.23}{1.70}
& \std{62.20}{1.49}
& \std{0.89}{0.58}
& \std{1.74}{2.43}
& \std{58.36}{1.71} \\

DARE
& 2024
& \std{59.39}{3.31}
& \std{8.70}{3.48}
& \std{-8.47}{3.68}
& \std{58.63}{1.33}
& \std{62.05}{1.74}
& \std{2.41}{2.63}
& \std{-0.26}{2.99}
& \std{60.42}{1.72} \\

DUCT
& 2025
& \std{56.12}{1.23}
& \std{8.57}{1.56}
& \std{-8.46}{1.55}
& \std{54.79}{0.48}
& \std{57.78}{2.57}
& \std{7.21}{2.35}
& \std{-2.46}{2.07}
& \std{53.58}{2.33} \\

BiLoRA
& 2025
& \std{56.12}{1.15}
& \std{5.11}{1.45}
& \std{-3.80}{1.65}
& \std{55.05}{1.33}
& \std{56.89}{1.86}
& \std{0.91}{0.37}
& \std{2.44}{1.36}
& \std{54.78}{1.17} \\

SOYO
& 2025
& \std{61.28}{1.43}
& \std{\underline{0.81}}{0.30}
& \std{-0.56}{0.36}
& \std{\underline{60.79}}{0.40}
& \std{64.68}{1.94}
& \std{\underline{0.67}}{0.30}
& \std{0.34}{0.61}
& \std{\textbf{62.29}}{1.06} \\

ContLoRA
& 2026
& \std{53.19}{1.28}
& \std{3.91}{2.20}
& \std{\textbf{1.77}}{1.75}
& \std{55.36}{0.52}
& \std{57.47}{1.20}
& \std{2.06}{0.24}
& \std{\underline{3.44}}{1.06}
& \std{57.70}{0.45} \\

Lung-CL
& 2026
& \std{62.20}{1.41}
& \std{2.41}{0.81}
& \std{\underline{0.68}}{1.76}
& \std{58.31}{0.31}
& \std{57.88}{0.91}
& \std{0.67}{0.65}
& \std{\textbf{3.55}}{1.95}
& \std{58.28}{0.74} \\

\midrule

Seq.~FT
& \textemdash
& \std{54.55}{1.60}
& \std{15.63}{3.04}
& \std{-15.63}{3.04}
& \std{56.98}{1.34}
& \std{64.99}{3.35}
& \std{4.44}{2.91}
& \std{-4.17}{3.43}
& \std{59.38}{2.06} \\

Joint Train.
& \textemdash
& \std{63.92}{0.81}
& \textemdash
& \textemdash
& \std{62.33}{0.64}
& \std{63.92}{0.81}
& \textemdash
& \textemdash
& \std{62.33}{0.64} \\

\midrule

\rowcolor{gray!12}
\textbf{APRIL}
& \textemdash
& \std{\textbf{64.43}}{0.85}
& \std{\textbf{0.37}}{0.14}
& \std{-0.32}{0.25}
& \std{\textbf{61.39}}{0.78}
& \std{\textbf{65.41}}{0.56}
& \std{\textbf{0.33}}{0.32}
& \std{0.02}{0.32}
& \std{\underline{61.27}}{1.16} \\

\bottomrule
\end{tabularx}

\vspace{0.6\baselineskip}

\caption{Performance comparison on binary SPRSound under
forward and reversed temporal orders. Reporting conventions,
abbreviations, and ranking rules follow
Table~\ref{tab:icbhi_main_comparison}.}
\label{tab:sprsound_main_comparison}

\begin{tabularx}{\textwidth}{
    @{}l
    @{\hspace{3pt}}c
    *{8}{>{\centering\arraybackslash}X}
    @{}
}
\toprule
\multirow{2}{*}{\textbf{Method}}
& \multirow{2}{*}{\textbf{Year}}
& \multicolumn{4}{c}{\textbf{Forward order}}
& \multicolumn{4}{c}{\textbf{Reversed order}} \\
\cmidrule(lr){3-6}
\cmidrule(lr){7-10}
&
& \textbf{FSA} ($\uparrow$)
& \textbf{AF} ($\downarrow$)
& \textbf{BWT} ($\uparrow$)
& \textbf{OS} ($\uparrow$)
& \textbf{FSA} ($\uparrow$)
& \textbf{AF} ($\downarrow$)
& \textbf{BWT} ($\uparrow$)
& \textbf{OS} ($\uparrow$) \\
\midrule

LwF
& 2016
& \std{\underline{82.99}}{1.08}
& \std{1.44}{1.14}
& \std{0.82}{1.62}
& \std{\underline{79.80}}{0.58}
& \std{80.49}{0.92}
& \std{4.59}{1.90}
& \std{-4.27}{1.88}
& \std{77.59}{0.65} \\

EWC
& 2017
& \std{77.61}{1.74}
& \std{7.89}{1.96}
& \std{-7.87}{1.99}
& \std{74.91}{1.70}
& \std{79.58}{0.69}
& \std{5.56}{1.23}
& \std{-5.13}{1.15}
& \std{77.25}{0.46} \\

ER
& 2019
& \std{80.97}{2.18}
& \std{4.01}{2.47}
& \std{-3.06}{3.03}
& \std{79.21}{2.04}
& \std{80.52}{0.95}
& \std{3.94}{1.42}
& \std{-3.66}{1.47}
& \std{77.80}{0.78} \\

DER++
& 2020
& \std{80.37}{1.67}
& \std{4.46}{2.10}
& \std{-4.12}{1.99}
& \std{78.23}{1.98}
& \std{81.56}{1.30}
& \std{2.72}{1.80}
& \std{-2.19}{1.48}
& \std{\underline{78.93}}{1.26} \\

InfLoRA
& 2024
& \std{80.80}{0.77}
& \std{2.82}{0.89}
& \std{-1.92}{0.81}
& \std{77.65}{0.84}
& \std{\underline{82.18}}{0.67}
& \std{0.97}{0.65}
& \std{0.07}{0.24}
& \std{78.90}{0.62} \\

DARE
& 2024
& \std{81.31}{1.62}
& \std{1.66}{0.98}
& \std{\underline{2.58}}{2.95}
& \std{79.27}{1.58}
& \std{81.51}{0.41}
& \std{\underline{0.58}}{0.74}
& \std{\textbf{2.29}}{0.56}
& \std{78.82}{0.38} \\

DUCT
& 2025
& \std{70.32}{2.24}
& \std{12.78}{2.45}
& \std{-12.33}{2.75}
& \std{68.48}{2.22}
& \std{73.97}{2.70}
& \std{7.38}{3.49}
& \std{-5.40}{3.22}
& \std{73.01}{1.88} \\

BiLoRA
& 2025
& \std{77.74}{0.31}
& \std{1.55}{0.27}
& \std{-0.64}{0.46}
& \std{74.75}{0.38}
& \std{78.22}{0.35}
& \std{1.83}{0.49}
& \std{-1.70}{0.70}
& \std{75.52}{0.49} \\

SOYO
& 2025
& \std{80.07}{0.99}
& \std{2.29}{0.69}
& \std{-1.49}{1.10}
& \std{79.07}{0.47}
& \std{79.41}{0.21}
& \std{0.95}{0.21}
& \std{-0.82}{0.23}
& \std{78.78}{0.24} \\

ContLoRA
& 2026
& \std{81.10}{0.48}
& \std{1.25}{0.57}
& \std{1.40}{0.89}
& \std{79.60}{0.30}
& \std{78.54}{1.04}
& \std{1.31}{0.72}
& \std{0.14}{0.88}
& \std{75.81}{0.91} \\

Lung-CL
& 2026
& \std{81.00}{0.33}
& \std{\textbf{0.36}}{0.22}
& \std{\textbf{3.23}}{0.84}
& \std{78.03}{0.34}
& \std{80.26}{1.01}
& \std{1.44}{0.18}
& \std{\underline{2.25}}{0.29}
& \std{76.88}{0.89} \\

\midrule

Seq.~FT
& \textemdash
& \std{76.96}{1.57}
& \std{8.62}{1.69}
& \std{-8.35}{2.06}
& \std{74.36}{1.67}
& \std{78.58}{0.10}
& \std{6.19}{0.30}
& \std{-5.61}{0.90}
& \std{76.41}{0.33} \\

Joint Train.
& \textemdash
& \std{83.25}{0.51}
& \textemdash
& \textemdash
& \std{81.79}{0.30}
& \std{83.25}{0.51}
& \textemdash
& \textemdash
& \std{81.79}{0.30} \\

\midrule

\rowcolor{gray!12}
\textbf{APRIL}
& \textemdash
& \std{\textbf{83.12}}{0.32}
& \std{\underline{0.81}}{0.29}
& \std{1.19}{0.19}
& \std{\textbf{80.34}}{0.30}
& \std{\textbf{82.23}}{0.91}
& \std{\textbf{0.52}}{0.30}
& \std{-0.08}{0.37}
& \std{\textbf{79.59}}{0.84} \\

\bottomrule
\end{tabularx}

\end{table*}

\begin{figure*}[t]
    \centering
    \includegraphics[width=0.90\textwidth]{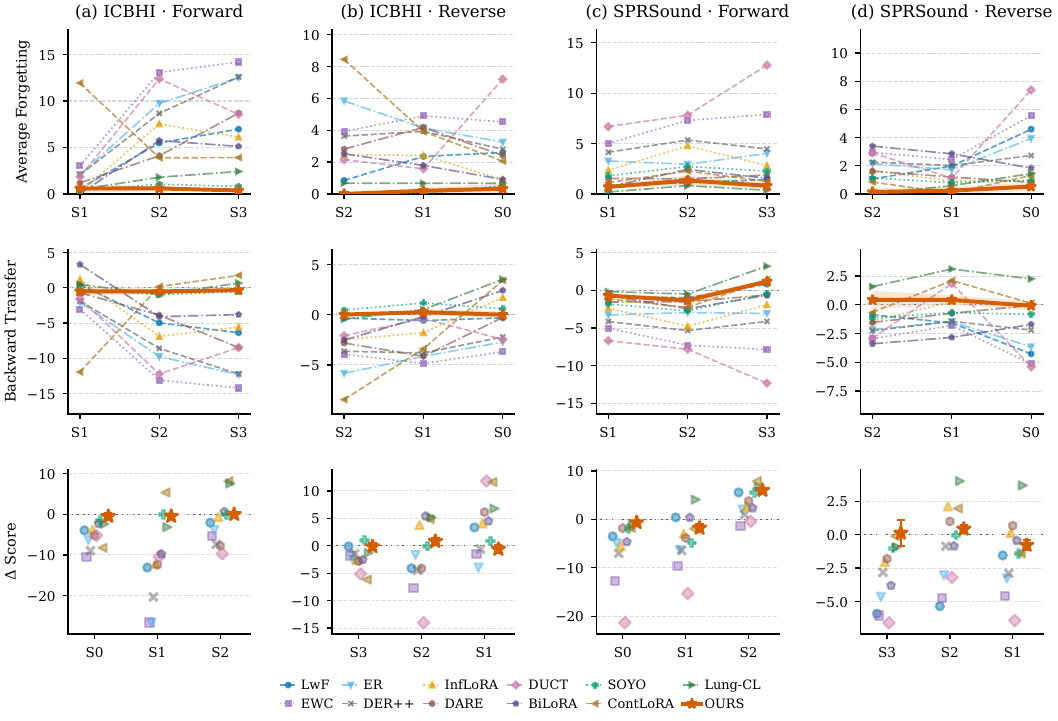}
   \caption{Forgetting dynamics on ICBHI and SPRSound. Columns (a)--(d) show ICBHI forward/reverse and SPRSound forward/reverse, respectively. Rows report stage-wise AF, BWT, and final domain-wise changes $\Delta\mathrm{Score}_{j}=R_{T-1,j}-R_{j,j}$. Values are five-seed means; shading and error bars show one standard deviation for APRIL.}
    \label{fig:forgetting_dynamics}
\end{figure*}

Tables~\ref{tab:icbhi_main_comparison} and
\ref{tab:sprsound_main_comparison} compare APRIL with
the selected continual-learning methods under forward
and reversed domain orders. APRIL achieves the highest
FSA across all four protocols, together with low
forgetting and competitive OS. Compared with Seq.~FT,
it improves all four metrics, showing that its higher
final classification performance is accompanied by
better retention of earlier domains.

On ICBHI, APRIL improves forward-order FSA by 1.85
points over the strongest competing method, LwF,
while keeping AF below 0.4 under both orders.
These results are important when considered together:
low forgetting alone does not indicate how well a
method learns incoming domains, while high final
performance can coexist with substantial losses on
earlier ones. APRIL combines strong final domain-wise
performance with limited degradation.
Under the reversed order, it retains the highest FSA,
although its OS ranks second to SOYO.

On SPRSound, APRIL also leads in FSA under both
orders and achieves the highest OS, extending its
effectiveness from device-defined domains to annual
releases. Its FSA margins over the strongest baselines
are smaller than on ICBHI, but its consistently low
forgetting strengthens the overall result.
The order comparisons further show that APRIL's
performance is not tied to a single sequence:
reversing either dataset changes its FSA by less
than one percentage point.

Some competing methods achieve higher positive BWT
in individual protocols. However, BWT measures how
much performance changes after a domain is first
learned, rather than how well that domain is
ultimately classified. A larger improvement from
a lower initial Score need not produce a higher
final Score. APRIL's near-zero or positive BWT,
combined with its leading FSA, indicates that it
maintains strong classification performance without
substantial deterioration during subsequent learning.

Joint Training provides an offline reference with
access to all training domains and achieves higher
OS than APRIL. Nevertheless, APRIL obtains higher
FSA under both ICBHI orders. This contrast illustrates
why both metrics are informative: a model can perform
better on the combined test set without achieving
a higher average across equally weighted domains.
APRIL's results show that sequential learning can
retain competitive overall performance while
supporting strong performance across recording domains.

\subsection{Forgetting dynamics}
\label{sec:forgetting_dynamics}

Figure~\ref{fig:forgetting_dynamics} shows how performance
on earlier domains evolves as new domains are learned.
APRIL maintains low AF across all four protocols,
with five-seed means not exceeding 1.33 points.
In contrast, EWC and ER show substantial increases
in AF under the forward ICBHI order, while DUCT
exhibits pronounced late-stage increases under
both SPRSound orders.
APRIL's BWT stays close to zero on ICBHI and reversed
SPRSound. Under forward SPRSound, it recovers from
an intermediate decline and becomes positive after
the final update.
Together, these trajectories show that APRIL retains
earlier-domain performance while continuing to learn
from incoming recordings.

The bottom row shows how subsequent learning affects
each earlier domain.
For each domain, $\Delta\mathrm{Score}$ is the Score
measured after all domains have been learned minus
the Score measured immediately after training on
that domain.
Negative values indicate that later learning reduces
its Score, while positive values indicate improvement.
On ICBHI and reversed SPRSound, APRIL's points remain
close to zero, showing that performance on earlier
domains is largely maintained.
Forward SPRSound shows a different pattern:
the 2022 and 2023 domains experience small declines,
whereas the 2024 domain improves by 6.00 points
after the final update.
This improvement outweighs the losses on the other
two domains, resulting in a positive final BWT of 1.19.

\subsection{Expert isolation and initialization}

\begin{table*}[t]
\centering
\caption{Expert isolation and initialization ablation.
\emph{w/o isolation} sequentially updates a shared head and
LoRA adapter; other rows use isolated experts.
Method names specify head/LoRA initialization:
Fre. denotes fresh initialization (random head or standard
LoRA initialization), and Inh. denotes inheritance from
the preceding expert. Best means are bold.}
\label{tab:expert_initialization}

\footnotesize
\setlength{\tabcolsep}{4pt}
\renewcommand{\arraystretch}{1.10}

\begin{tabularx}{\textwidth}{
    @{}l
    *{8}{>{\centering\arraybackslash}X}
    @{}
}
\toprule
\multirow{2}{*}{\textbf{Method}}
& \multicolumn{4}{c}{\textbf{Forward order}}
& \multicolumn{4}{c}{\textbf{Reversed order}} \\
\cmidrule(lr){2-5}
\cmidrule(lr){6-9}
&
\textbf{FSA} ($\uparrow$)
& \textbf{AF} ($\downarrow$)
& \textbf{BWT} ($\uparrow$)
& \textbf{OS} ($\uparrow$)
& \textbf{FSA} ($\uparrow$)
& \textbf{AF} ($\downarrow$)
& \textbf{BWT} ($\uparrow$)
& \textbf{OS} ($\uparrow$) \\
\midrule

\multicolumn{9}{@{}l}{\textbf{ICBHI}} \\

\emph{w/o isolation}
& \std{53.74}{2.09}
& \std{16.50}{3.27}
& \std{-16.35}{3.17}
& \std{59.15}{1.15}
& \std{62.62}{0.80}
& \std{4.56}{1.71}
& \std{-3.98}{1.52}
& \std{56.38}{0.84} \\

\cmidrule(lr){1-9}

Fre. / Fre.
& \std{63.03}{0.46}
& \std{0.97}{0.32}
& \std{-0.97}{0.32}
& \std{60.27}{0.71}
& \std{61.64}{0.63}
& \std{0.41}{0.19}
& \std{-0.34}{0.26}
& \std{59.14}{1.22} \\

Inh. / Fre.
& \std{63.08}{0.80}
& \std{0.96}{0.33}
& \std{-0.96}{0.33}
& \std{59.86}{0.81}
& \std{63.04}{0.61}
& \std{\textbf{0.29}}{0.09}
& \std{\textbf{0.21}}{0.27}
& \std{59.77}{1.39} \\

\rowcolor{gray!12}
Inh. / Inh.
& \std{\textbf{64.43}}{0.85}
& \std{\textbf{0.37}}{0.14}
& \std{\textbf{-0.32}}{0.25}
& \std{\textbf{61.39}}{0.78}
& \std{\textbf{65.41}}{0.56}
& \std{0.33}{0.32}
& \std{0.02}{0.32}
& \std{\textbf{61.27}}{1.16} \\

\midrule
\multicolumn{9}{@{}l}{\textbf{SPRSound}} \\

\emph{w/o isolation}
& \std{78.10}{0.69}
& \std{6.89}{0.98}
& \std{-6.41}{1.18}
& \std{76.62}{0.61}
& \std{79.04}{0.39}
& \std{4.76}{0.75}
& \std{-4.37}{0.35}
& \std{76.86}{0.29} \\

\cmidrule(lr){1-9}

Fre. / Fre.
& \std{82.67}{0.32}
& \std{0.97}{0.17}
& \std{0.94}{0.35}
& \std{79.78}{0.28}
& \std{\textbf{82.67}}{0.55}
& \std{\textbf{0.51}}{0.27}
& \std{-0.19}{0.24}
& \std{\textbf{79.84}}{0.47} \\

Inh. / Fre.
& \std{82.15}{0.42}
& \std{0.86}{0.30}
& \std{\textbf{1.40}}{0.35}
& \std{79.57}{0.40}
& \std{82.28}{0.66}
& \std{0.56}{0.08}
& \std{-0.11}{0.20}
& \std{79.79}{0.64} \\

\rowcolor{gray!12}
Inh. / Inh.
& \std{\textbf{83.12}}{0.32}
& \std{\textbf{0.81}}{0.29}
& \std{1.19}{0.19}
& \std{\textbf{80.34}}{0.30}
& \std{82.23}{0.91}
& \std{0.52}{0.30}
& \std{\textbf{-0.08}}{0.37}
& \std{79.59}{0.84} \\

\bottomrule
\end{tabularx}
\end{table*}

Table~\ref{tab:expert_initialization} examines expert isolation
and initialization across both datasets and domain orders.
We compare a shared head and LoRA adapter updated sequentially
with isolated experts using three initialization schemes:
fresh initialization of both components, inheritance of the
head alone, and inheritance of both the head and LoRA adapter.
Fresh initialization uses a randomly initialized head and
standard LoRA initialization; inherited initialization copies
the corresponding parameters from the preceding expert.

Using a shared head and LoRA adapter instead of isolated
experts leads to substantially greater forgetting across
all four protocols. Compared with APRIL, AF increases from
0.37 to 16.50 on forward ICBHI and from 0.81 to 6.89 on
forward SPRSound. Both reversed orders also show higher
AF and more negative BWT. The shared model repeatedly
updates the same parameters to learn successive domains,
whereas isolated experts allow new-domain adaptation
without modifying earlier experts. These results support
expert isolation as an effective design for reducing
forgetting during sequential learning.

Within the isolated-expert setting, inheriting the head alone
does not consistently improve final performance.
Inheriting both components improves FSA and OS over head-only
inheritance in three of the four protocols, suggesting that
transferring the adapter together with its associated head
provides a more useful starting point for adaptation.
The benefit is not universal: on reversed SPRSound,
fresh initialization achieves the highest FSA and OS,
while inheriting both components gives the highest BWT.
Thus, initialization mainly affects how effectively a new
expert adapts, and its benefit depends on the dataset and
domain order. APRIL retains joint head--LoRA inheritance
as a common configuration across protocols, rather than
claiming that it optimizes every metric in every setting.

\subsection{Routing and expert selection}

\label{sec:routing_analysis}
\begin{table*}[t]
\centering
\caption{Factorial ablation of routing features and expert selection on ICBHI and SPRSound under forward and reversed domain orders. Hard selects the top-ranked expert. Top-2 confidence selects between the two top-ranked experts using routing and classifier confidence.}
\label{tab:routing_factorial_four_protocols}
\footnotesize
\setlength{\tabcolsep}{4pt}
\renewcommand{\arraystretch}{1.10}

\begin{tabularx}{\textwidth}{
    @{}llll
    *{4}{>{\centering\arraybackslash}X}
    @{}
}
\toprule
Dataset & Order & Routing features & Expert selection & FSA ($\uparrow$) & AF ($\downarrow$) & BWT ($\uparrow$) & OS ($\uparrow$) \\
\midrule
\multirow{8}{*}{ICBHI}
 & \multirow{4}{*}{Forward} & Final layer & Hard & \std{64.09}{0.92} & \std{0.39}{0.20} & \std{-0.32}{0.29} & \std{61.02}{0.91} \\
 & & Multi-level & Hard & \std{\textbf{64.48}}{0.83} & \std{\textbf{0.29}}{0.11} & \std{\textbf{-0.23}}{0.21} & \std{\textbf{61.41}}{0.79} \\
 & & Final layer & Top-2 confidence & \std{64.11}{0.92} & \std{0.39}{0.20} & \std{-0.32}{0.29} & \std{61.09}{0.94} \\
 & & \cellcolor{gray!12}Multi-level
& \cellcolor{gray!12}Top-2 confidence
& \cellcolor{gray!12}\std{64.43}{0.85}
& \cellcolor{gray!12}\std{0.37}{0.14}
& \cellcolor{gray!12}\std{-0.32}{0.25}
& \cellcolor{gray!12}\std{61.39}{0.78} \\
\cmidrule(lr){2-8}
 & \multirow{4}{*}{Reverse} & Final layer & Hard & \std{65.17}{0.38} & \std{\textbf{0.18}}{0.10} & \std{0.43}{0.24} & \std{61.16}{1.02} \\
 & & Multi-level & Hard & \std{\textbf{65.43}}{0.55} & \std{0.29}{0.31} & \std{0.09}{0.36} & \std{\textbf{61.28}}{1.16} \\
 & & Final layer & Top-2 confidence & \std{65.21}{0.36} & \std{0.19}{0.11} & \std{\textbf{0.49}}{0.32} & \std{61.14}{1.05} \\
 & & \cellcolor{gray!12}Multi-level
& \cellcolor{gray!12}Top-2 confidence
& \cellcolor{gray!12}\std{65.41}{0.56}
& \cellcolor{gray!12}\std{0.33}{0.32}
& \cellcolor{gray!12}\std{0.02}{0.32}
& \cellcolor{gray!12}\std{61.27}{1.16} \\
\midrule
\multirow{8}{*}{SPRSound}
 & \multirow{4}{*}{Forward} & Final layer & Hard & \std{82.00}{0.47} & \std{1.39}{0.27} & \std{0.29}{0.45} & \std{79.85}{0.34} \\
 & & Multi-level & Hard & \std{82.12}{0.50} & \std{1.34}{0.32} & \std{0.40}{0.41} & \std{80.08}{0.31} \\
 & & Final layer & Top-2 confidence & \std{82.44}{0.52} & \std{1.13}{0.27} & \std{0.74}{0.41} & \std{80.02}{0.34} \\
 & & \cellcolor{gray!12}Multi-level
& \cellcolor{gray!12}Top-2 confidence
& \cellcolor{gray!12}\std{\textbf{83.12}}{0.32}
& \cellcolor{gray!12}\std{\textbf{0.81}}{0.29}
& \cellcolor{gray!12}\std{\textbf{1.19}}{0.19}
& \cellcolor{gray!12}\std{\textbf{80.34}}{0.30} \\
\cmidrule(lr){2-8}
 & \multirow{4}{*}{Reverse} & Final layer & Hard & \std{80.84}{0.95} & \std{0.88}{0.29} & \std{-0.31}{0.28} & \std{78.86}{0.85} \\
 & & Multi-level & Hard & \std{80.98}{0.82} & \std{0.90}{0.28} & \std{-0.54}{0.16} & \std{79.14}{0.73} \\
 & & Final layer & Top-2 confidence & \std{81.41}{0.97} & \std{0.79}{0.35} & \std{-0.33}{0.29} & \std{79.05}{0.85} \\
 & & \cellcolor{gray!12}Multi-level
& \cellcolor{gray!12}Top-2 confidence
& \cellcolor{gray!12}\std{\textbf{82.23}}{0.91}
& \cellcolor{gray!12}\std{\textbf{0.52}}{0.30}
& \cellcolor{gray!12}\std{\textbf{-0.08}}{0.37}
& \cellcolor{gray!12}\std{\textbf{79.59}}{0.84} \\
\bottomrule
\end{tabularx}
\end{table*}

\begin{table*}[t]
\centering
\caption{Routing diagnostics.
Top-1 accuracy and Top-2 recall indicate whether the true
domain is retrieved first or among two candidates.
Domain-oracle uses the corresponding domain expert.
Gap is domain-oracle FSA minus routed FSA, computed per seed.
FSA and Gap are mean $\pm$ standard deviation over five seeds.
Domain labels are used only for diagnostics.}
\label{tab:routing_diagnostics}
\small
\setlength{\tabcolsep}{7pt}
\renewcommand{\arraystretch}{1.10}
\begin{tabular}{llccccc}
\toprule
Dataset & Order & Top-1 Acc. (\%) & Top-2 Recall (\%) &
Domain-oracle FSA $\uparrow$ & Routed FSA $\uparrow$ & Gap $\downarrow$ \\
\midrule
\multirow{2}{*}{ICBHI}
& Forward & 83.60 & 93.53 & \std{65.16}{0.74} & \std{64.43}{0.85} & \std{0.72}{0.45} \\
& Reverse & 83.49 & 93.27 & \std{65.81}{0.50} & \std{65.41}{0.56} & \std{0.41}{0.23} \\
\midrule
\multirow{2}{*}{SPRSound}
& Forward & 55.98 & 83.66 & \std{82.92}{0.39} & \std{83.12}{0.32} & \std{-0.20}{0.18} \\
& Reverse & 55.90 & 83.63 & \std{82.13}{0.43} & \std{82.23}{0.91} & \std{-0.11}{0.79} \\
\bottomrule
\end{tabular}
\end{table*}

\begin{figure*}[t]
\centering
\includegraphics[width=0.90\textwidth]{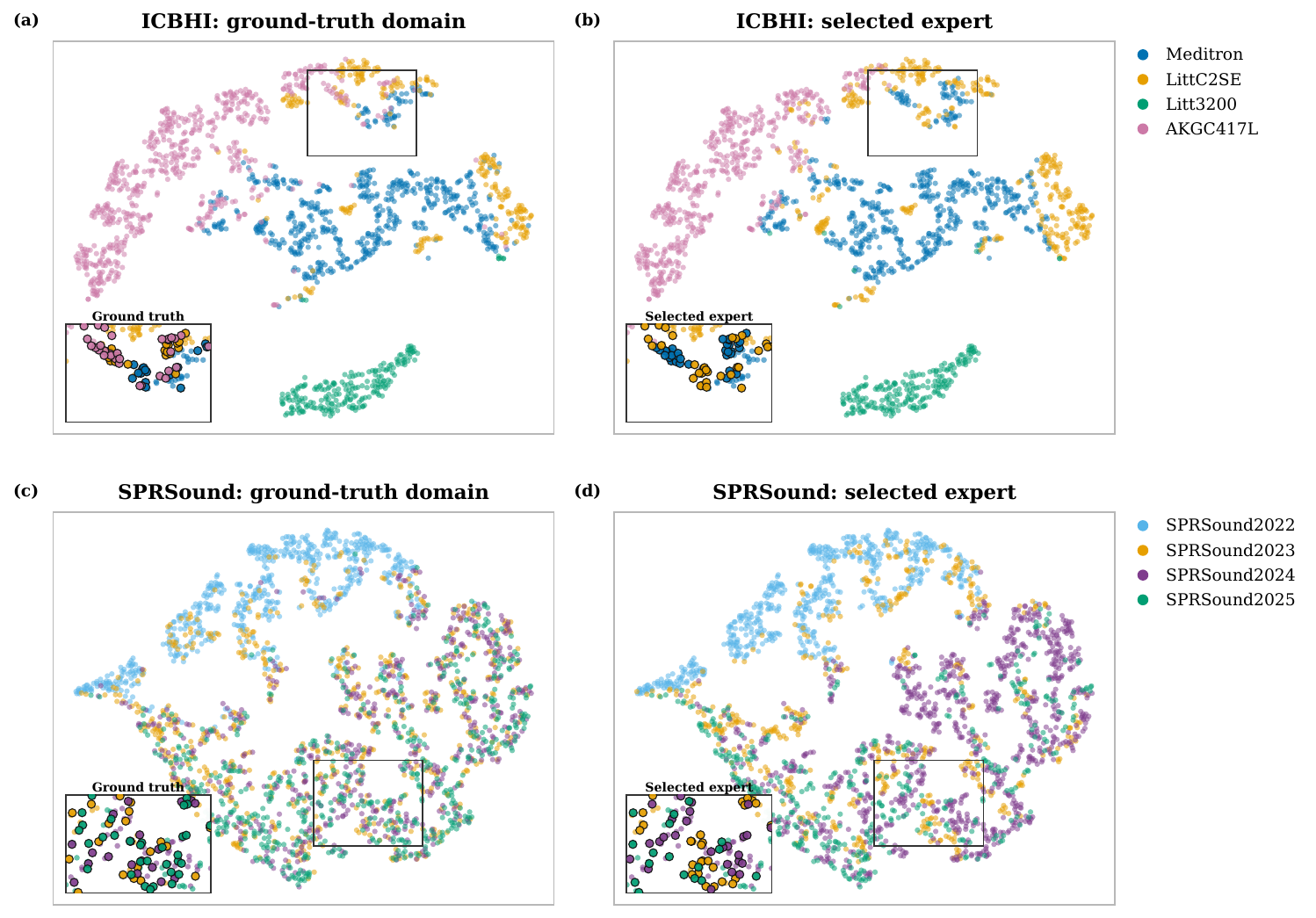}
\caption{t-SNE visualization of multi-level domain-matching
scores under the forward order. The same recordings are
colored by their true domains (left) and selected experts
(right). Paired insets magnify the same region; black outlines
mark recordings assigned to an expert from a different domain.}
\label{fig:routing_tsne}
\end{figure*}

Table~\ref{tab:routing_factorial_four_protocols} compares
routing features and expert-selection strategies while
keeping the trained expert banks fixed and using two
prototypes per domain at each feature level.
Multi-level features improve mean FSA and OS over
final-layer features under both selection rules across
all four protocols. On SPRSound, these improvements
are larger with Top-2 confidence selection than with
hard selection, suggesting that information from
different encoder depths is particularly useful when
combined with a comparison of candidate predictions.

The benefit of Top-2 confidence selection differs
between the datasets. With multi-level features,
it improves all four metrics on SPRSound, increasing
FSA by 1.00 and 1.25 points under the forward and
reversed orders, respectively.
On ICBHI, the differences are much smaller, and
hard selection gives slightly higher FSA and lower AF.
Evaluating a second expert therefore provides a clear
benefit on SPRSound, but contributes little additional
improvement on ICBHI.

The retrieval results in
Table~\ref{tab:routing_diagnostics} help explain this
difference. Top-1 domain accuracy is approximately
84\% on ICBHI but 56\% on SPRSound, while Top-2 recall
reaches approximately 94\% and 84\%, respectively.
For many SPRSound recordings, the expert trained on
their own domain is therefore retrieved as the second
candidate rather than the first.
Top-2 selection makes this expert available for
comparison instead of committing immediately to
the highest-ranked domain.

Figure~\ref{fig:routing_tsne} provides a visual
explanation of the retrieval results.
The left column colors recordings by their actual
domains, while the right column colors the same
recordings by their selected experts.
On ICBHI, recordings from different devices form
relatively distinct groups, and most retain similar
colors in the selected-expert panel.
On SPRSound, the 2022 recordings form a more distinct
group, whereas recordings from 2023--2025 overlap
extensively. The enlarged regions highlight recordings
assigned to experts trained on another year's data.
The mixture of annual domains makes domain matching
less straightforward, helping explain why comparing
two candidate experts is more useful on SPRSound.

Table~\ref{tab:routing_diagnostics} also compares
APRIL with domain-oracle assignment, which uses
the expert trained on each recording's own domain.
APRIL's FSA remains within one point of this
reference on ICBHI under both orders and is slightly
higher on SPRSound.
All experts recognize the same respiratory sound
classes, so an expert trained on another domain
can still classify a recording correctly.
Experts from different domains may also provide
complementary class-specific knowledge
~\cite{li2025addressing}.
These results explain why APRIL can maintain strong
classification performance on SPRSound despite
its lower domain-retrieval accuracy.

\begin{table}[t]
\centering
\caption{Deployment cost of APRIL using Top-2 confidence selection.
Latency is measured with a batch size of 32 and reported over three
repetitions. Storage excludes the shared frozen backbone.}
\label{tab:efficiency}
\footnotesize
\setlength{\tabcolsep}{5.5pt}
\renewcommand{\arraystretch}{1.10}
\begin{tabular}{@{}lccc@{}}
\toprule
Dataset
& \shortstack{Bank params.\\(M)}
& \shortstack{Storage\\(MiB)}
& \shortstack{Latency\\(ms/sample)} \\
\midrule
ICBHI
& 2.372 & 9.12 & \std{4.81}{0.01} \\
SPRSound
& 2.365 & 9.09 & \std{4.97}{0.04} \\
\bottomrule
\end{tabular}
\end{table}

\subsection{Sensitivity to prototype count}
\label{sec:parameter_sensitivity}

\begin{figure*}[t]
    \centering
    \includegraphics[width=0.90\textwidth]{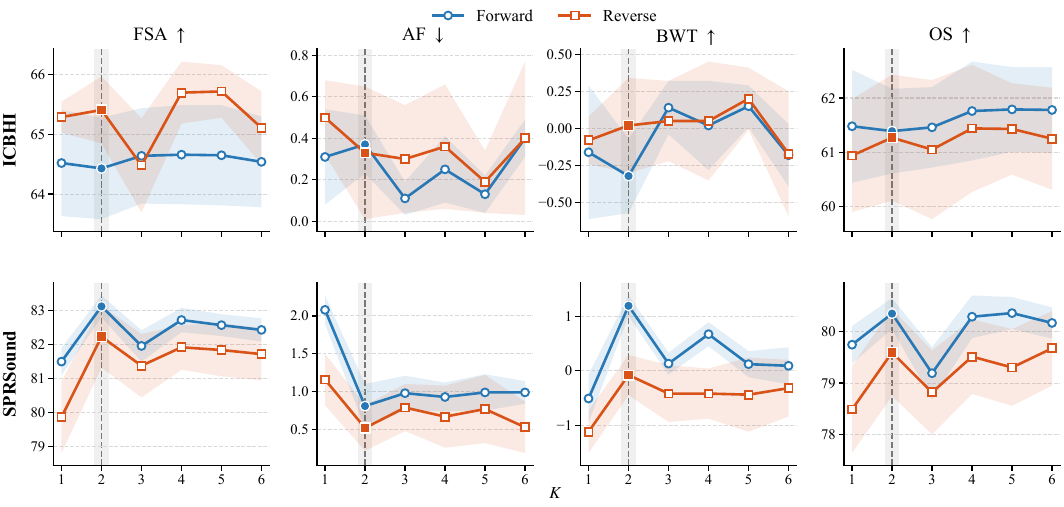}
    \caption{Sensitivity to prototype count $K$.
    Curves and shading show five-seed means and standard
    deviations.}
    \label{fig:k_sensitivity}
\end{figure*}

Figure~\ref{fig:k_sensitivity} examines the number of
prototypes per domain while fixing the expert bank and
other routing components.
On ICBHI, $K=1$ already provides competitive performance:
its FSA is only 0.14 and 0.43 points below the best
results under the forward and reversed orders,
respectively.
Nevertheless, $K=4$ or $K=5$ improves several metrics.
Although recordings from different devices are relatively
well separated in the reference space, those acquired
using the same device still vary across patients and
respiratory sound classes.
Multiple prototypes may capture this variation better
than a single center.

SPRSound benefits more substantially from increasing
$K$ from one to two, with FSA gains of 1.62 and 2.36
points under the forward and reversed orders,
respectively, together with lower AF and higher OS.
Further increases, however, do not consistently improve
performance.
Because the annual domains share overlapping acoustic
characteristics, representing each domain more finely
does not necessarily make the domains easier to distinguish.
The results therefore suggest that prototype count
depends on both within-domain variation and between-domain
overlap, rather than domain separation alone.
We adopt $K=2$ across all protocols: it achieves the best
SPRSound FSA, AF, and BWT under both orders while maintaining
competitive ICBHI performance with fewer stored prototypes
than $K=4$ or $K=5$.

\subsection{Storage and computational cost}

Table~\ref{tab:efficiency} summarizes the deployment cost
of APRIL. Each new domain adds approximately 0.59M
parameters while reusing the shared frozen BEATs backbone.
For four domains, the expert bank and routing prototypes
occupy 9.12 MiB on ICBHI and 9.09 MiB on SPRSound,
excluding the shared backbone.
This compact design accommodates new devices or acquisition
periods without storing a separate full model for each domain.
During inference, APRIL evaluates only two retrieved experts,
with measured latencies of 4.81 and 4.97 ms per sample
on ICBHI and SPRSound, respectively.
By combining small domain-specific parameter additions
with a fixed number of candidate expert evaluations,
APRIL offers a scalable and privacy-friendly deployment paradigm for lifelong respiratory sound monitoring.

\section{Discussion}
\label{sec:discussion}

This study investigates continual respiratory sound classification
when recordings arrive from successive domains and historical data
are unavailable for replay. APRIL combines sequentially inherited
LoRA--classifier experts with multi-level prototype retrieval and
ambiguity-aware expert selection. Across the four evaluation
protocols, APRIL achieves the highest FSA among the compared
continual-learning methods, reaching 64.43 and 65.41 on ICBHI
and 83.12 and 82.23 on SPRSound under the forward and reversed
orders, respectively. It also obtains the lowest AF in three
protocols. Beyond these overall results, the ablations highlight
the complementary roles of expert preservation and selection.
Keeping earlier experts unchanged substantially reduces forgetting,
while the routing design determines how effectively their knowledge
is used when domain identities are unavailable. The benefits of
confidence-guided selection are more pronounced on SPRSound,
where recordings from different annual releases show greater
overlap in the reference space. Its strong classification performance,
despite lower domain-retrieval accuracy, further suggests that
identifying the recording's domain and selecting a useful classifier
are not equivalent objectives. These findings support combining
domain relevance with classification evidence when selecting
experts for respiratory recordings acquired under evolving conditions.

Several limitations motivate further investigation.
The evaluation covers two benchmarks with four predefined domains
each and fixed label spaces. Longer domain sequences and more
diverse recording conditions are needed to assess how well the
observed adaptation and retention extend beyond these protocols.
Although APRIL evaluates only two candidate experts per recording,
its expert bank grows with every additional domain.
Future work will explore expert consolidation and adapter
compression to control storage growth while preserving
domain-specific knowledge. Moreover, the reported runtime reflects
batched GPU inference, whereas portable respiratory assessment
may require low-latency processing on resource-constrained hardware.
Hardware-specific optimization and prospective evaluation in
clinical workflows will therefore be important next steps toward
practical use.

\FloatBarrier
\bibliographystyle{unsrt}
\bibliography{cas-refs}
\end{document}